# Two mechanisms of formation of asthenospheric layers

L. Czechowski and M. Grad

Institute of Geophysics, Faculty of Physics, University of Warsaw
Ul. Pasteura 5, 02-093 Warszawa, Poland
Phone: +48 22 5532003
E-mail: lczech@op.pl

Corresponding author:
Leszek Czechowski: lczech@op.pl

The theory of plate tectonics describes some basic global tectonic processes as a result of motion of lithospheric plates. The boundary between lithosphere and asthenosphere (LAB) is defined by a difference in response to stress. Position of LAB is determined by: (i) the ratio (melting temperature)/( temperature) and (ii) an invariant of the stress tensor. We consider the role of these both factors for origin and decay of asthenosphere. We find that the asthenosphere of shear stress origin could be a transient, time-dependent feature.

1. **Introduction**

The plate tectonics theory describes some of the basic tectonic processes on the Earth as motion of lithospheric plates. The plates are moved by large-scale thermal convection in the mantle. The lithosphere is mechanically resistant. Its thickness varies from ~50 to ~250 km. The lithosphere is underlain by the asthenosphere. The boundary between the lithosphere and the asthenosphere (LAB) is defined by a difference in response to stress: the asthenosphere deforms viscously. The bulk of asthenosphere is not melted [1], but at the deformation rate typical for the mantle convection (about $10^{-14}$ s$^{-1}$) it behaves as a fluid with the viscosity $\eta$ of about $5 \cdot 10^{19}$ kg m$^{-1}$ s$^{-1}$. The mantle below has higher effective viscosity (e.g. $10^{22}$ kg m$^{-1}$ s$^{-1}$) [2]. In the present paper we consider the role of two important factors determining the rheology: the ratio $q$=(melting temperature)/(temperature) and $\sigma$=(invariant of the stress tensor) for the evolution of the system of two lithospheric plates and two asthenospheres.

2. **Viscosity of the mantle**

A few factors determine effective viscosity of the mantle. According to [3,4,5] a generalized viscosity relation valid for both diffusion and dislocation creeps is:

$$\eta = C\left(\frac{\sigma}{G}\right)^{1-n} \exp\left(\frac{E_a + pV_a}{RT}\right), \qquad (1)$$

where $\eta$ is the viscosity [Pa s], $\sigma$ is the differential stress [Pa] (in fact: an invariant of the stress tensor), $E_a$ is the energy activation [J], $p$ is the pressure [Pa], and $V_a$ is the volume activation [m$^3$]. $C\,(G, h, f_{OH}, f_{melts})$ is the function of rigidity $G$ [Pa], the grain size $h$ [m], $f_{OH}$ is the hydrogen and hydroxyl concentration, $f_{melts}$ is the melt fraction [4].

The expression $E_a+pV_a$ is proportional to the temperature of melting $T_m$. In a typical approach, the asthenosphere is believed to be determined by the ratio of $T_m$ to the actual temperature $T$, i.e. by $q$. Therefore, the basic differences of lithosphere and asthenosphere properties are often explained as a result of the $T$-$p$ conditions.

The lithosphere is a thermal boundary layer [6] for the mantle convection. The temperature of its upper part is low but the vertical temperature gradient in the lithosphere is high. Below the lithosphere, the temperature gradient is low (could be close to the adiabatic gradient [7 8]) and $T_m(p)$ is increasing with depth faster than $T$. Hence, $q$ and the viscosity $\eta$ reach their minimum values just below LAB and they are increasing with depth in the mantle below.

Of course, it is a typical situation. Tectonic processes in subduction zones could change this picture. A lithospheric plate could be placed in the mantle below another plate. Distribution of $q$ in such a case could have two minima, so two asthenospheric layers could be formed.

Note however that equation (1) indicates that the effective viscosity depends also on other factors, e.g. chemical composition, size of grains, etc. In our consideration we choose to investigate the role of $\sigma$. Generally, viscosity is proportional to $\sigma^{1-n}$. For $n=1$ the viscosity does not depend on the stress (this is the Newtonian rheology). For true mantle $n$ is probably in the range from 3 to 5.

The asthenosphere plays an important role in the plate tectonics enabling the motion of plates. Note that the term "*asthenosphere*" could be understood in a few ways. Below a moving lithospheric plate, through the asthenosphere, there is a vertical gradient of horizontal velocity of the matter. This asthenosphere could be referred as an *active asthenosphere*. If there is no velocity gradient, but the viscosity is low, then one can use term: *potential asthenosphere*. It is a low viscosity and low velocity of seismic P- and S-waves zone (LVZ). The flow of the mantle material in the asthenosphere could result in some seismic anisotropy of the layer. The anisotropy could be permanent ('frozen') and consequently a *fossil seismic asthenosphere* could be formed.

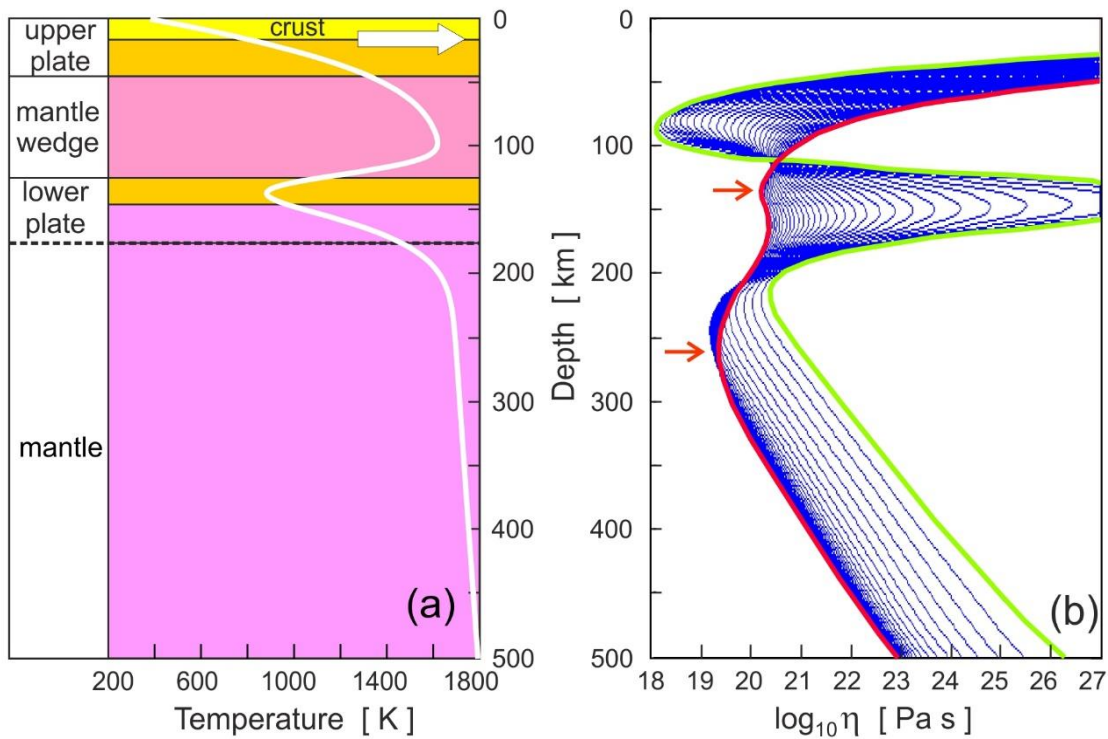

Fig. 1. (a): The scheme of our system and positions of the plates. White line gives the initial temperature. The upper plate's velocity is $10^{-9}$ m s$^{-1}$ (~3 cm yr$^{-1}$). (b): Evolution of the effective viscosity of the system for non-Newtonian rheology ($n=3.5$). The initial profile is green while the final is red (after 50 My). Blue lines give profiles for equal time steps (~2.5 My). Note that the final profile has two distinct minima of viscosity shown by the arrows. See text for more explanations.

### 3. Equations of our model

We investigate here the role of $q$, $\sigma$ and $n$ for the formation of asthenospheric layers. The $\sigma$ and $n$ are not directly coupled with the temperature, so our research includes also non-thermal conditions of the formation. We neglects only some factors included in $C(G, h, f_{OH}, f_{melts})$.

The existence of thermal ($q$) and mechanical ($\sigma$) factors means that (at least in some cases)

one can discriminate an asthenosphere of mainly thermal origin (i.e. determined by the value of *q*) and an asthenosphere of mainly shear stresses origin (determined by the value of *σ* ). This idea is presented in [9] and in more developed form in [10].

Table 1
Values of parameters used in the models. The $E_a$, $V_a$, and C/G are after [6]. Lower part gives thicknesses of a considered layers and heat production per 1 m$^{-3}$. Velocity at the lower boundary is zero.

| *n* | $E_a$ [J] | $V_a$ [m$^3$ mole$^{-1}$] | C/G | Velocity at the surface [m s$^{-1}$] | *g(x)* [Pa m$^{-1}$] | *k* [W m$^{-1}$ K$^{-1}$] | *ρ* [kg m$^{-3}$] |
|---|---|---|---|---|---|---|---|
| 3.5 (non-Newtonian) | 5.4×10$^5$ | 2×10$^{-5}$ | 10$^{10}$ | 10$^{-9}$ | 0.1 | 2 | 3500 |

| **Structure of plates** | | | | | | | |
|---|---|---|---|---|---|---|---|
| | $Q_{granite}$ [W m$^{-3}$] | Thickness *d* [km] | $Q_{basalt}$ [W m$^{-3}$] | Thickness *d* [km] | $Q_{mantle}$ [W m$^{-3}$] | Thickness *d* [km] | Total thickness[km] |
| Upper plate | 5×10$^{-7}$ | 18 | 10$^{-7}$ | 28 | 10$^{-8}$ | 80 | 126 |
| Lower plate | | 0 | 2×10$^{-7}$ | 20 | 10$^{-9}$ | | 20 |

To investigate that problem we use 1D model based on the following system of Navier-Stokes equation for infinite Prandtl number and equation of heat transfer:

$$\frac{\partial}{\partial x} \eta \frac{\partial}{\partial x} v = g(x) \qquad (2)$$

$$c\rho \frac{\partial T}{\partial t} = k \frac{\partial^2}{\partial x^2} T + Q , \qquad (3)$$

where *v* is the horizontal velocity [m s$^{-1}$], *g(x)* is the horizontal gradient of the pressure [Pa m$^{-1}$], *c* is the specific heat for constant pressure [J kg$^{-1}$ K$^{-1}$], *ρ* is the density [kg m$^{-3}$], *t* is time [s], *k* is the coefficient of thermal conduction [W m$^{-1}$ K$^{-1}$] and *Q* is the heat generation rate per unit volume [W m$^{-3}$]. Our model could describe basic processes determining evolution and extinction of the asthenosphere but still is simple and numerically efficient.

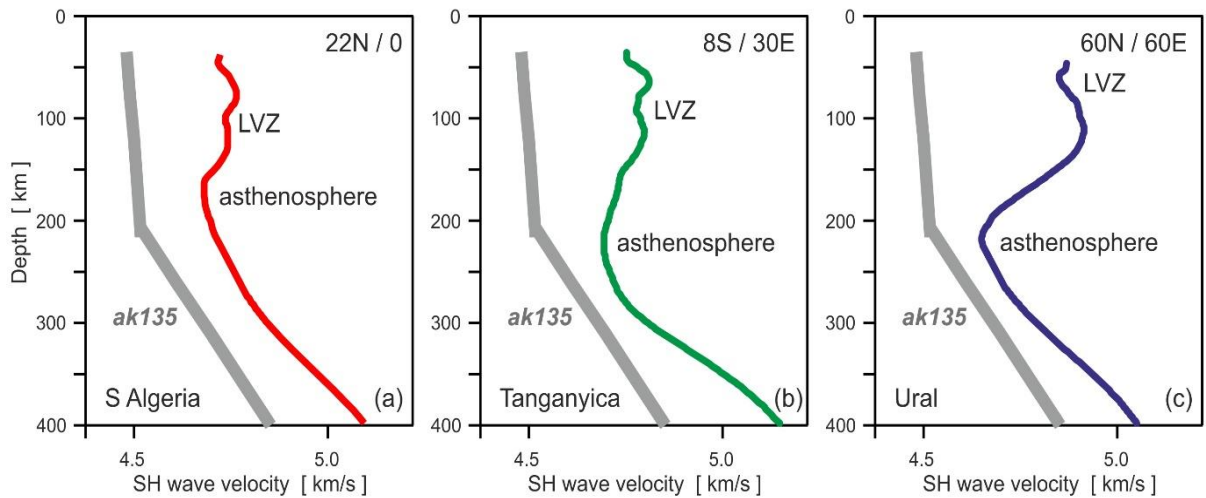

Fig.2. Seismological observations of low velocity zones and asthenosphere (according to data of [18 19] ). In many regions two layers with low velocity of S waves are observed. We suggest that at least some of LVZ could be presently `active`, `potential` or `fossil` asthenospheres.

4. **Simulations and results**

We consider two lithospheric plates: the upper plate and the lower plate (Fig. 1). The upper plate is a model of continental lithosphere which consists of thick crust and the lithospheric mantle. The lower plate is a model of oceanic lithosphere. The basaltic crust is thinner and its radiogenic heating is lower than in the upper plate – Fig. 1a and Table 1. Such system could serve as an approximation of the situation observed in some subduction zones, especially for flat-slab subduction [11,12,13]. The flat slab can extend for 100-1000 kilometers. The flat-slab subduction is used as an explanation of segmentation of the Andean Volcanic Belt.

The initial situation is given in Fig. 1a and in Table 1. White line in the Fig. 1a gives the initial temperature. According to our boundary conditions the temperature on the upper boundary is constant, and the heat flow (or temperature gradient) is constant on the lower boundary. The initial temperature profile corresponds to thermal age of the upper lithosphere of 30 My and for the lower plate: 20 My. For oceanic lithosphere the thermal age results from the cooling plate model [5]. For continental lithosphere it is just simplification. We do not model any specific region, but some parts of the continental lithosphere in Andean are geologically relatively young (i.e., ~50-100 My). The oceanic Nazca plate close to South America is also young, i.e. ~20-40 My.

Fig. 1b presents evolution of the viscosity distribution of our system for non-Newtonian rheology. The initial profile of viscosity is given by the green line while the final by the red one. Blue lines give velocity profiles for equal time step, approximately each 2.5 My. Note that the final profile has two distinct minima shown by arrows. The upper minimum could be treated as an expression of a `fossil asthenosphere` and the lower one as an `active asthenoshere`. For the Newtonian reology the minima of viscosity are less distinct.

Note an important difference between the evolution of $T$-$p$ conditions and mechanical conditions included in $\sigma$. The typical time scale $\tau$ [s] for thermal evolution is given by [15]:

$$\tau = \lambda^2 / \kappa, \qquad (4)$$

where $\lambda$ is the spatial scale [m] (e.g. the thickness of asthenosphere) and $\kappa = k/(\rho c)$ is thermal diffusivity [m$^2$ s$^{-1}$]. For typical values: $\lambda$=100 km and $\kappa$=10$^{-6}$ m$^2$ s$^{-1}$ the $\tau$=300 My. So the characteristic time of decay of the thermal asthenosphere is hundreds of My.

The evolution of mechanical conditions could be faster. Consider moving an oceanic lithospheric plate with a terrane. When the terrane reaches the subduction zone the resistance is high and could even stop the plate in geologically short time. The corresponding changes of $\sigma$ could be fast and consequently the characteristic time of decay of the asthenosphere of the shear stress origin could be short.

5. **Other options and observations**

Our model concerns mainly flat-slab subduction [11]. However, a few other processes could also lead to double asthenospheric layers. An example of such process is given in [14]. His Fig. 3 presents model for continent-continent collision with lithospheric delamination (stage 4 in the figure). Another example is given by [16] (his Fig. 6.16).

In Fig. 2 we give some profiles of the seismic wave velocities from different regions. Two minima of the velocity is a quite common phenomenon. Compare the minima of viscosity with minima of the velocity of SH waves (LVZ and asthenosphere).

We suggest that some of them could be a result of the existence of double asthenospheric layers. Seismic tomographic observation presented in [17] gives other examples of complicated structures of a few low and high velocity layers. Other problems with determination of LAB are discussed in [6,13].

6. **Conclusions**

1. Two kinds of asthenosphere could exist: mainly of mechanical origin (or shear stress origin) and mainly of thermal origin.
2. The evolution of $\sigma$ could be very fast, so the asthenosphere of shear stress origin could be a transient, time-dependent feature.
3. The evolution of the system of two asthenospheres leads usually to origin of a thick lithosphere

with two low S-wave velocity layers. Note also that two low S-wave velocity layers are observed in some continental lithospheres.

**Acknowledgments**
This work was partially supported by the National Science Centre (grant 2011/01/B/ST10/06653). Computer resources of Interdisciplinary Centre for Mathematical and Computational Modeling of University of Warsaw were also used in the research

**References**
1. Olugboji T.M., Karato, S. and Park, J. (2013), Structures of the oceanic lithosphere-asthenosphere boundary: Mineralphysics modeling and seismological signatures, *Geochem. Geophys. Geosyst.,* **14**, 880–901, doi:10.1002/ggge.20086.
2. Harig, C., Zhong, S., and Simons, F.J. 2010. Constraints on upper mantle viscosity from the flow induced pressure gradient across the Australian continental keel. *Geochem. Geophys. Geosyst*., **11**, Q06004, doi:10.1029/2010GC003038.
3. Shintaro A., Katayama I., Nakakuki, T. 2014. Rheological decoupling at the Moho and implication to Venusian tectonics. *Nature*. **4** : 4403 | DOI: 10.1038/srep04403.
4. Kohlstedt, D.L., 2006. Water and rock deformation: The case for and against a climb-controlled creep rate, in Water in Nominally Anhydrous Minerals, eds. H. Keppler and J.R. Smyth, *Reviews in Mineralogy and Geochemistry, Mineralogical Society of America*, **62**, 377-396. (2006)
5. Turcotte, D. L., and Schubert, G.. 2002. Geodynamics, Cambridge Academic Press, pp. 456.
6. Hamza V. M. and Vieira, F. P., 2012, Global distribution of the lithosphere-asthenosphere boundary: a new look. *Solid Earth*, **3**, 199–212.
7. Putirka, K.D., Perfit, M., Ryerson, F.J., Jackson, M. G.. 2007. Ambient and excess mantle temperatures, olivine thermometry, and active vs. passive upwelling. *Chemical Geology* **241**, 177–206.
8. da Silva Cesar R.S., Wentzcovitch, R.M., Patel, A., Price, G.D., Karato, S.I., 2000. The composition and geotherm of the lower mantle: constraints from the elasticity of silicate perovskite. *Physics of the Earth and Planetary Interiors* **118**, 103–109.
9. Czechowski, L., 2012. Origin of LAB as a result of dynamical and thermal processes. 4$^{th}$ PASSEQ 2006-2008 Workshop in Jachranka, Poland, 14-17 of March, (2012).
10. Czechowski, L. and Grad, M. 2013. Origin and evolution of asthenospheric layers. Geophysical Research Abstracts, Vol. **15**, EGU2013-8951.
11. Gutscher, M-A., Spakman, W., Bijwaard, H., Engdahl, R., 2000. Geodynamics of flat subduction: Seismicity and tomographic constraints from the Andean margin. *Tectonics*, 19, 5, 814-833.
12. English, J.M., Johnston, S.T., Wang, K. 2003. Thermal modelling of the Laramide orogeny: testing the £at-slab subduction hypothesis *Earth Planet. Sci. Let*. **214**, 619-632.
13. Eaton, D.W., Darbyshire, F., Evans, R.L., Grutter, H., Jones, A.G., Yuan, X., 2009. The elusive lithosphere-asthenosphere boundary (LAB) beneath cratons. *Lithos* **109**, 1-22.
14. Massonne H.-J., 2005. Involvement of Crustal Material in Delamination of the Lithosphere after Continent-Continent Collision, *International Geology Review*, **47**, 792–804.
15. Czechowski, L. (1993) Theoretical approach to mantle convection in Teisseyre, R., Czechowski, L., Leliwa-Kopystynski, J. (eds.), Dynamics of the Earth's evolution, Elsevier-PWN, p. 161-271.
16. Stiiwe K., 2006. Geodynamics of the Lithosphere: An Introduction, Springer.
17. Liu, K., Levander, A., Zhai, Y., Porritt, R.W., Allen, R.M., 2012. Asthenospheric flow and lithospheric evolution near the Mendocino Triple Junction, *Earth Planet Sci. Lett,* 323-324, 60-71.
18. Ritzwoller, M.H., Shapiro, N.M., Barmin, M.P., and Levshin, A.L. Global surface wave diffraction tomography, *J. Geophys. Res.*, **107** (B12), 2335.
19. Shapiro, N.M., Ritzwoller, M.H., 2002. Monte-Carlo inversion for a global shear velocity model of the crust and upper mantle. Geophysical Journal International **151**, 8–105. http://ciei.colorado.edu/~nshapiro/MODEL/.